\documentclass[preprint,aps,showpacs,preprintnumbers,amsmath,amssymb]{revtex4}

\usepackage{graphicx}% Include figure files
\usepackage{dcolumn}% Align table columns on decimal point
\usepackage{bm}% bold math

\begin{document}

%Title of paper
\title{Artifacts with uneven sampling of red noise}
\author{Edoardo Milotti}
\email{milotti@ts.infn.it}
\affiliation{Dipartimento di Fisica, Universit\`a di Trieste and I.N.F.N. -- Sezione di Trieste\\
Via Valerio, 2 -- I-34127 Trieste, Italy}

\date{\today}

\begin{abstract}
The vast majority of sampling systems operate in a standard way: at each tick of a fixed-frequency master clock a digitizer reads out a voltage that corresponds to the value of some physical quantity and translates it into a bit pattern that is either transmitted, stored, or processed right away. Thus signal sampling at evenly spaced time intervals is the rule: however this is not always the case, and uneven sampling is sometimes unavoidable. 

While periodic or quasi-periodic uneven sampling of a deterministic signal can reasonably be expected to produce artifacts, it is much less obvious that the same happens with noise: here I show that this is indeed the case only for long-memory noise processes, i.e., power-law noises $1/f^\alpha$ with $\alpha > 2$. The resulting artifacts are usually a nuisance although they can be eliminated with a proper processing of the signal samples, but they could also be turned to advantage and used to encode information. 
\end{abstract}

\maketitle

\section{Introduction}
\label{intro}

Nearly all digital signal-measuring equipment found in laboratories throughout the world, from the humble voltmeter up to powerful computerized data-logging systems and high-frequency digitizing scopes, operate with an internal master clock that sets the pace for an analog-to-digital converter that translates the electrical output of a transducer into a bit pattern. The master clock frequency is usually held as stable as possible and the sample intervals are fixed to a very high degree of precision. Correspondingly, most signal-analysis techniques are meant to be used on evenly spaced data: this is true for the Discrete Fourier Transform (DFT) and also for Autoregressive (AR) or Moving Average (MA) modeling of data \cite{KM}. However some data happen to be unevenly sampled: this is especially true for astronomers, who are seldom so lucky as to have an uninterrupted series of clear nights, and in general are bound to observe whatever comes from the sky, whenever it comes, and have to search for periodicities amid these scattered data. Indeed it was the astronomical community that developed the first effective spectral estimation techniques for unevenly sampled data \cite{LS}. 

Uneven sampling has special properties: Beutler proved rigorously \cite{Beu} that in general uneven sampling is not band-limited and later showed that a random, Poisson-distributed set of sampling times allows perfect signal reconstruction \cite{Beu2}. Earlier, Yen \cite{Yen} was able to derive modified forms of the Shannon reconstruction formula for different types of uneven sampling, which are however much more complex than the corresponding formula for even sampling. And indeed, randomly sampled signals are not easy to analyze and many standard methods must be abandoned, although in some cases one can restore regular sampling using reconstruction algorithms \cite{Vio}. 

Since no sampling clock is quite perfect and is normally affected by noise \cite{Barnes} and by deterministic drifts (that may be periodic), all regular sampling should actually be regarded as quasi-regular sampling. Ignoring this may be dangerous, because it is clear that quasi-regular sampling of a deterministic signal (e.g. a sinusoidal signal) may introduce unwanted harmonics in the DFT analysis of the sampled signal, unless corrective measures are taken. But what happens if one samples pure noise? Can one still produce artifacts? If this were the case then an unrecognized quasi-regular sampling of a noisy background might become a problem since it would produce fake signals that could be mistaken for true. In this paper I show that this is just what happens in some cases of colored noise, i.e., in the case of the  long-memory noise processes $1/f^\alpha$ with $\alpha > 2$, while the correlation between samples for noises with $\alpha \le 2$ is insufficient to produce replicas of the low-frequency peak of the noise spectrum. A proof is given in section \ref{proof1}, while section \ref{sim} illustrates numerical results that confirm the theoretical analysis, and section \ref{disc} discusses some implications of  these findings.

\section{Proof based on a DFT model of the noise process}
\label{proof1}

It is well known that a signal sampled $N$ times in the time interval $(0,T)$ can be modeled by a sum of $N$ exponentials and that this is equivalent to a DFT: in other words the DFT is a physical model of the signal \cite{KM} and we can write: 
\begin{equation}
\label{DFT}
f_n=\frac{1}{\sqrt{N}}\sum_{k=0}^{N-1}F_k \exp\left(\frac{2\pi i nk} {N}\right)
\end{equation}
where $f_n$ denotes the $n$-th sample, and the fit coefficients $F_k$ correspond to the DFT and can be calculated from the formula 
\begin{equation}
\label{InvDFT}
F_k=\frac{1}{\sqrt{N}}\sum_{n=0}^{N-1}f_n \exp\left(-\frac{2\pi i nk} {N}\right).
\end{equation}
With the usual regular sampling intervals $\Delta t$, the $n$-th sampling time is $t_n=n \Delta t$, the total sampling time is $T=N \Delta t$, and equation (\ref{DFT}) can be rewritten as follows: 
\begin{equation}
\label{DFT2}
f_n=\frac{1}{\sqrt{N}}\sum_{k=0}^{N-1}F_k  \exp\left(2\pi i k\frac{t_n} {T}\right)
\end{equation}
If sampling is not quite regular, the sampling times $t_n$ are replaced by $t_n+\Delta t_n$, and equation (\ref{DFT2}) becomes
\begin{equation}
\label{DFT2uneven}
f_n=\frac{1}{\sqrt{N}}\sum_{k=0}^{N-1}F_k  \exp\left(\frac{2\pi i nk} {N} + 2\pi i \frac{k}{N}\frac{\Delta t_n} {\Delta t}\right)
\end{equation}
Now we assume a periodic sampling pattern with a period equal to $M$ clock ticks, so that $N = m M$ and $M \ll N$; we also assume at first that $m$ is an integer (numerical simulations show that this requirement can be relaxed and $m$ can be real) and we expand the relative timing shift $\Delta t_n/\Delta t$ as a Fourier sum: 
\begin{equation}
\label{rts}
\frac{\Delta t_n} {\Delta t}=\frac{1}{\sqrt{M}}\sum_{l=0}^{M-1}\phi_l \exp\left(\frac{2\pi i n m l} {N}\right)
\end{equation}
and in addition, we assume the relative timing shifts to be very small, i.e., $\Delta t_n/\Delta t \ll 1$. Next we notice that $k/N<1$, and then, using equations (\ref{DFT2uneven}) and (\ref{rts}) and after a few cumbersome but straightforward passages, we can approximate the observed DFT with the following formula: 
\begin{equation}
\label{obsDFT}
F'_k \approx F_k + \frac{2\pi i}{N\sqrt{M}}\sum_{l=0}^{M-1}(k-ml)\phi_l F_{k-ml}
\end{equation}
moreover if we make the rather weak assumption that the phase of the noise DFT in different frequency bins is uncorrelated so that $\langle F_k F_l\rangle = 0$ if $k \neq l$ where $\langle \rangle$ is the usual ensemble average, then the DFT  (\ref{obsDFT})  gives the following spectrum:
\begin{eqnarray}
\nonumber
S'_k & = & \frac{\langle \left| F'_k \right|^2 \rangle}{N} = \\
\nonumber
& = & \frac{1}{N} \left\langle |F_k|^2 + \frac{2\pi i}{N\sqrt{M}}\sum_{l=0}^{M-1}(k-ml)(\phi_l F^*_k F_{k-ml} - \phi^*_l F_k F^*_{k-ml}) \right. \\
\nonumber
&& + \left. \frac{4\pi^2}{N^2 M}\sum_{l,l'=0}^{M-1} (k-ml)(k-ml')\phi^*_l \phi_l' F^*_{k-ml} F_{k-ml'} \right\rangle \\
\label{obsPSD}
& \approx & S_k + \frac{4\pi^2}{N^2 M}\sum_{l=0}^{M-1}(k-ml)^2 \left|\phi_l\right|^2 S_{k-ml}
\end{eqnarray}
(the hypothesis of phase independence is quite common, because it is essential for noise generators like that of Timmer and K\"onig \cite{TK}, and is supported by the numerical results reported in \cite{milotti}).

If the noise is white, i.e., the spectral density is flat, or if it is a $1/f^\alpha$ noise  with a spectral index $\alpha \leq 2$, we see from eq. (\ref{obsPSD}) that the periodic uneven sampling amounts to the addition of a (small) non-flat background. In fact a $1/f^\alpha$ noise has a discrete spectrum $S_k \approx C/k^\alpha$, therefore the observed spectrum (\ref{obsPSD}) becomes 
\begin{equation}
S'_k \approx S_k + \frac{4\pi^2 C}{N^2 M}\sum_{l=0}^{M-1}(k-ml)^{2-\alpha} \left|\phi_l\right|^2 
\end{equation}
However, if the spectral index $\alpha$ is greater than 2, then the $l$-th harmonic of the relative time shift $\Delta t_n/\Delta t$ produces a peak over the power-law background, which is just  the low-frequency noise peak, shifted to the $(m l)$-th frequency bin.

\section{Numerical simulation}
\label{sim}

The analysis that leads to eq. (\ref{obsPSD}) assumes small relative timing shifts, but in this section I report numerical simulations carried out with the exact power-law noise generator described in \cite{milotti,milotti2,milotti3} that do support the analytical results also for large relative timing shifts \cite{note}. 

The generator used in the simulation runs produces power-law noise from a superposition of random exponential pulses, and is exact in the sense that it produces a process that is theoretically guaranteed to yield a range-limited power-law spectrum between two extreme (angular) frequencies $\lambda_{min}$ and $\lambda_{max}$. The generator takes correctly into account the correlation between samples in colored noises, and works also with unevenly spaced sampling times. 

\begin{figure}
\includegraphics[width=5in]{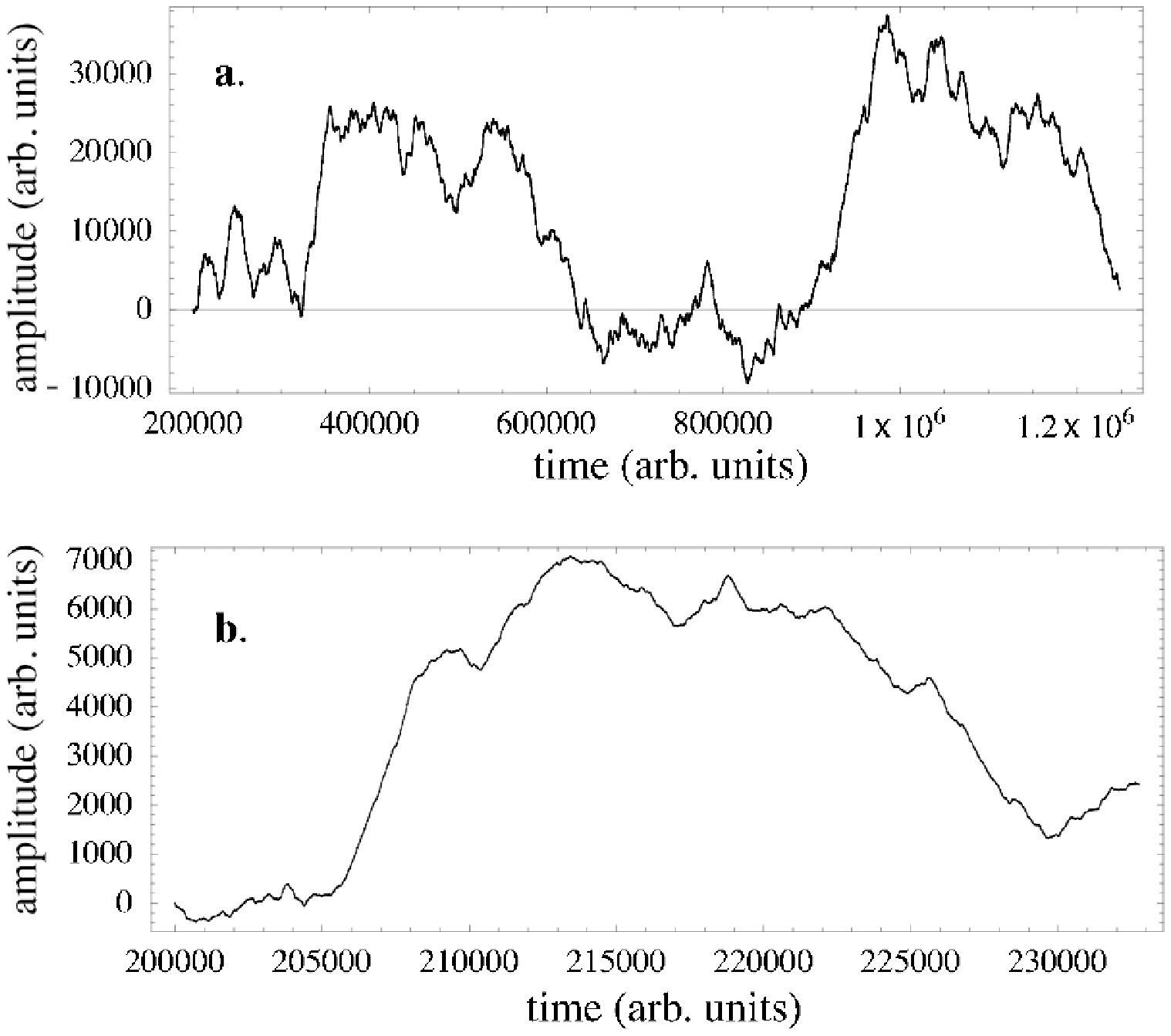}
\caption{\label{fig1} This figure show a noise signal produced with the generator described in \cite{milotti,milotti2,milotti3}. The parameters in this specific run are $\alpha=3$, $\lambda_{min} = 0.0001$, and $\lambda_{max}=1$, i.e., the spectrum has a power-law region $1/f^3$ that spans the angular frequency interval $\lambda_{min}  < \omega < \lambda_{max}$. The generator produces power-law noise from a superposition of random exponential pulses, and in this run the pulse rate has been set at $n = 10$ pulses per unit time, so that the resulting noise signal is Gaussian to a very high degree \cite{milotti}. Time is in arbitrary units, and the average sampling interval is $\Delta t = 1$ (arb. units); the choice of time units also sets the corresponding frequency units used for the relaxation rates $\lambda_{min} $ and $\lambda_{max}$. The sampling time has been sinusoidally modulated: $\Delta t_k/\Delta t = 1+ 0.2 \sin(2\pi k/4)$, and the signal has been sampled $2^{20} = 1048576$ times. Part {\bf a}. shows the whole signal generated in this run (time does not start from zero, because at the beginning some samples are used for the generator initialization and are discarded); part {\bf b}. shows the initial 32768 valid samples. Notice that even though the sampling time modulation is rather large (20\%), it is quite invisible in the zoomed figure.
}
\end{figure}

\begin{figure}
\includegraphics[width=5in]{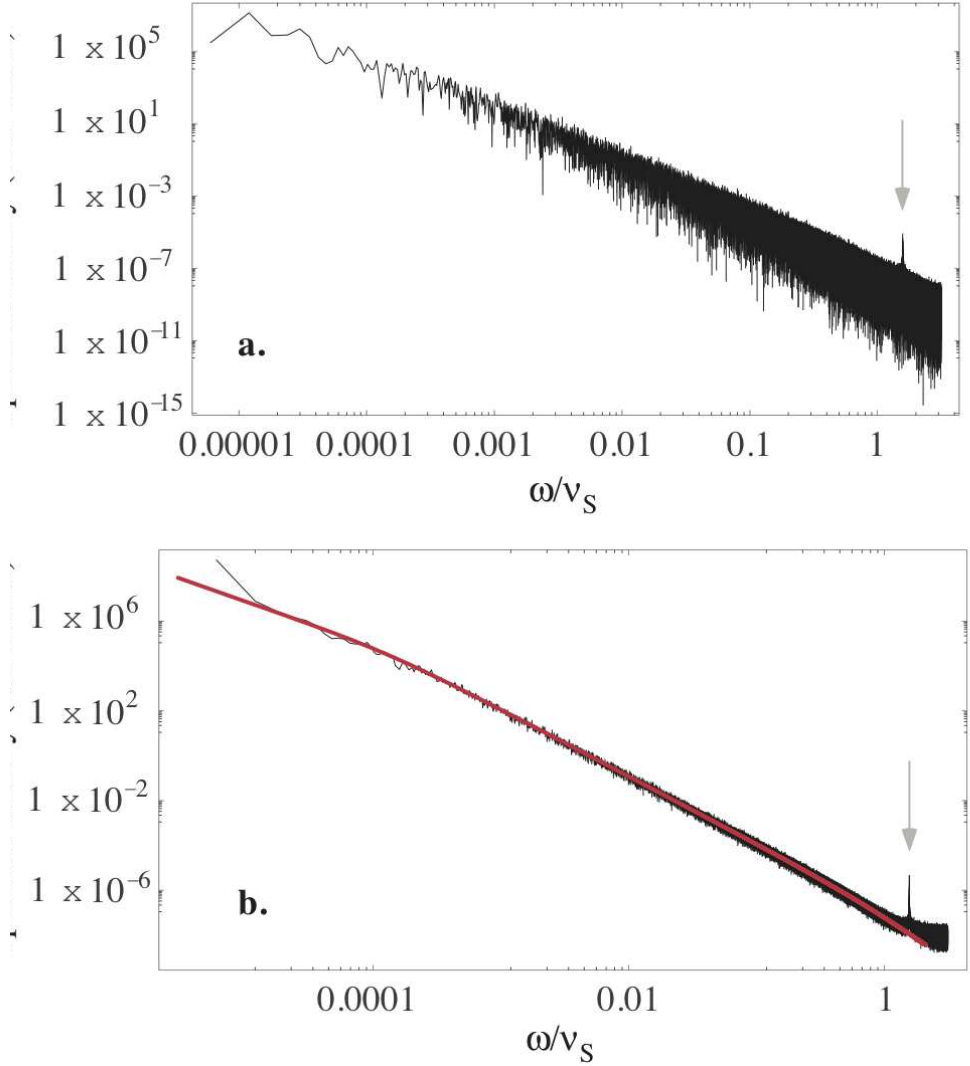}
\caption{\label{fig2} {\bf a.} Spectrum of the signal shown in figure (\ref{fig1}) vs. the scaled angular frequency $\omega/\nu_S$. The arrow marks the peak from sampling time modulation at the expected position $\omega = \omega_N/2$, where $\omega_N=\pi \nu_S$ is the (angular) Nyquist frequency and $\nu_S$ is the sampling frequency. {\bf b.} Spectrum averaged over 16 realizations of the same noise process: the solid line shows the expected (theoretical) behavior of the noise spectrum \cite{milotti2,milotti3}, corrected for the incoherent gain of the Hanning window that has been used for trend removal. The upward bend at high frequency in the spectrum {\bf b} is due to aliasing which becomes apparent after averaging, while the upward bend at low frequency is due to the uncorrected DC component which cannot be removed by windowing. }
\end{figure}

In these simulations, time is in arbitrary units, and the average sampling interval is $\Delta t = 1$ arb. units; the choice of time units also sets the corresponding frequency units used for the relaxation rates $\lambda_{min} $ and $\lambda_{max}$. 
Figure \ref{fig1} shows a simulated signal obtained with the noise generator for a $1/f^3$ noise; in this case the generator parameters are  $\alpha=3$, $\lambda_{min} = 0.0001$, and $\lambda_{max}=1$, i.e., the spectrum has a power-law region $1/f^3$ that spans the frequency interval $1.6\cdot 10^{-5}  < \omega < 1.6\cdot 10^{-1}$, and the pulse rate has been set at $n = 10$ pulses per unit time, so that the resulting noise signal is Gaussian to a very high degree \cite{milotti}. In this case the sampling time has been sinusoidally modulated: $\Delta t_k/\Delta t = 1+ 0.2 \sin(2\pi k/4)$ (the period for uneven sampling is 4 samples), and there are in all $2^{20} = 1048576$ samples. 
Figure \ref{fig2} shows the DFT spectrum of the signal of figure (\ref{fig1}): a comparison with the exact theoretical spectrum of the noise generator \cite{milotti2,milotti3}
\begin{eqnarray}
\nonumber
S(\omega) & = & \frac{1}{(\lambda_{max}^{1-\beta} - \lambda_{min}^{1-\beta})\omega^4}\left[  \lambda_{max}^{1-\beta} F\left(\frac{1-\beta}{2},1;\frac{1-\beta}{2};\frac{-\lambda_{max}^2}{\omega^2}\right) \right. \\
\label{psdplaw}
& & \left.-\lambda_{min}^{1-\beta} F\left(\frac{1-\beta}{2},1;\frac{1-\beta}{2};\frac{-\lambda_{min}^2}{\omega^2}\right)  \right]
\end{eqnarray}
-- which has a $1/f^\alpha$ power-law in the range $\lambda_{min} < \omega < \lambda_{max}$ -- shows that on the whole the sampled noise process produced by the noise generator actually behaves as predicted by theory \cite{milotti,milotti2,milotti3}, except for a small peak at the frequency of the sampling time modulation. This small peak only shows up in this and in other runs (not shown here, but easily reproducible \cite{note}) with $\alpha > 2$, and this lends support to the proofs of the previous sections. The spectra shown in this and in the other figures have been partly detrended with a Hanning window (a general introduction to the need of the detrending step can be found, e.g., in \cite{xu}; see also the qualitative considerations in \cite{mandel,schroeder}).

A closer look at the modulation peaks yields however a much more striking confirmation of the analytical results: in fact the theoretical spectral density (\ref{psdplaw}) of the noise generator has a $1/f^\alpha$ power-law region for $\lambda_{min} < \omega < \lambda_{max}$, while for  $\omega < \lambda_{min}$ it has a $1/f^2$ behavior, and this means that from equation (\ref{obsPSD}) we expect that the correction term due to uneven sampling is negligible just at the modulation frequency, while there should be two side-peaks whose exact shape depends on the low-frequency limit of the $1/f^\alpha$ region, i.e., on $\lambda_{min}$. And indeed this is just what happens in the simulations, as shown in figure \ref{fig3}, where part {\bf a.} shows the region of the averaged spectrum in figure \ref{fig2}b close to the peak due to sampling time modulation. Figure \ref{fig3}b shows the expected behavior calculated from equation (\ref{obsPSD}) and from the conditions used in the generation of the noise process and listed above (the correction for the incoherent gain of the Hanning window is also included). Finally, figure \ref{fig3}c is the superposition of parts \ref{fig3}a and \ref{fig3}c, where we see that the calculated shape closely matches the observed shape.

Figure \ref{fig4} shows the results of a similar calculation performed on the spectrum of a single realization of the noise process shown in figure \ref{fig2}a: figure \ref{fig4}a is the zoomed portion of the spectrum around the modulation peak. Figure \ref{fig4}b is a still closer zoom, and the arrow in the figure shows the position of the modulation frequency: notice that there is no peak just where one would naively expect to find one. The low frequency part of the spectrum in figure \ref{fig2}a has been inserted in equation (\ref{obsPSD}) to obtain the spectrum in figure \ref{fig4}c, and we see that there is an almost perfect correspondence with the peak in figure \ref{fig4}c: this means that the correlation terms between different frequencies (the cross-terms in the derivation of equation (\ref{obsPSD})) are negligible even for a single realization of the noise process.

I have also noted that there must be a dependence of the split-peak shape on the exact shape of the low-frequency part of the spectrum: figure \ref{fig5}a shows the shape of the peak for a larger value of $\lambda_{min}$ ($\lambda_{min}=10^{-3}$). In this case the low-frequency part of the noise spectrum (\ref{psdplaw}) has a wider $1/f^2$ region, and the side-peaks must be correspondingly lower and further apart: indeed this is just what happens in figure \ref{fig5}a. A further confirmation is provided by figure \ref{fig5}b, which shows the peak for a smaller value of $\lambda_{min}$ ($\lambda_{min}=5\cdot 10^{-5}$): the side-peaks are much higher and also closer.

\section{Discussion}
\label{disc}

While most of the observed power-law ($1/f^\alpha$) noises have spectral indexes  $0 < \alpha \leq 2$, with an apparent clustering around $\alpha = 1$, red noises, i.e., noises with spectral indexes $\alpha > 2$, also show up in several unrelated systems \cite{mandel,schroeder} like the water level of the Nile river, economics, orchid population size \cite{gillman} and local temperature fluctuations and affect precise timekeeping \cite{vessot} and our ability to predict environmental and animal population variables \cite{kim}. Noises with $\alpha > 2$ also appear in the energy level fluctuations of quantum systems \cite{rel,sal} and in timing noise in pulsars \cite{scott}. Because of their extreme peaking behavior at low frequencies  these noises are also called ``black'' \cite{schroeder}, and they display marked persistence properties \cite{mandel} that may lead to the mistaken identification of underlying trends in experimental data \cite{rang}. From the results reported in this paper it follows that these noises pose yet another potential danger to experiments that use uneven sampling, because their long-memory properties give rise to artifacts in the DFT spectra.

\begin{figure}
\includegraphics[width=4in]{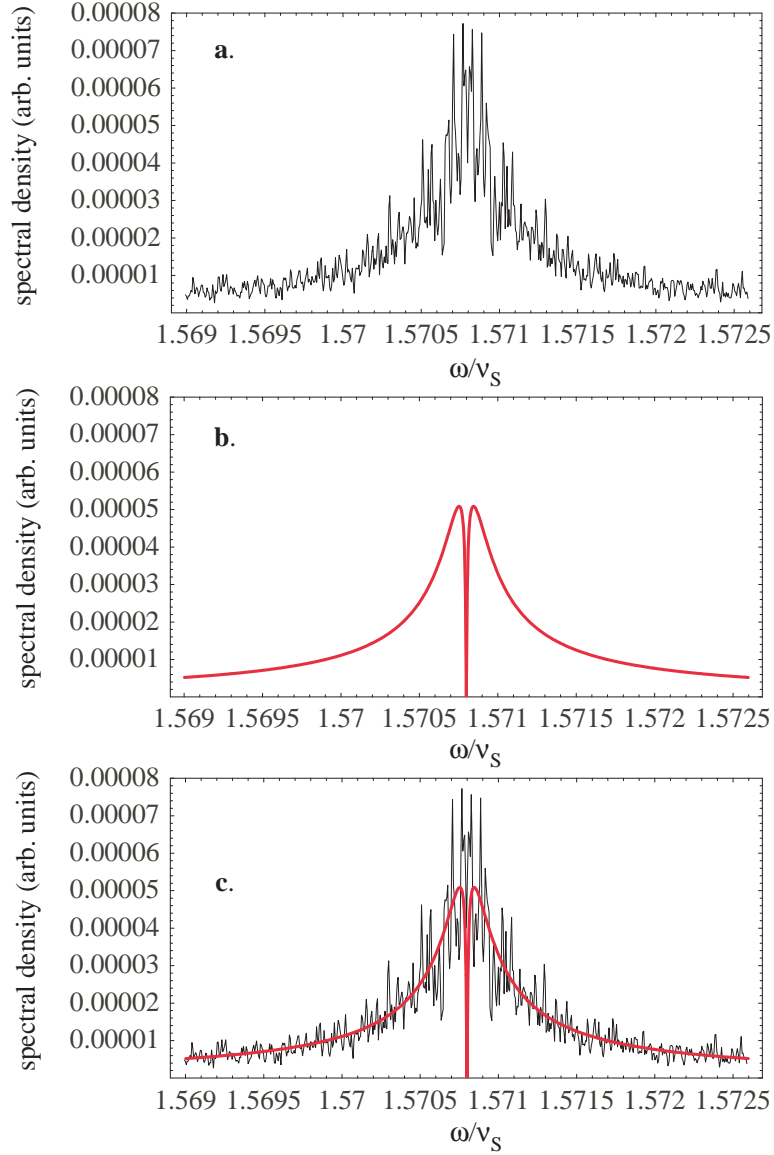}
\caption{\label{fig3} {\bf a}.) Averaged spectrum of figure \ref{fig2}b zoomed and centered on the position of the small peak (i.e. $\omega/\nu_S = \pi/2$), with linear scales on both axes. {\bf b}.) Expected behavior calculated from equation (\ref{obsPSD}) and from the conditions used in the generation of the signal in figure \ref{fig1} (the correction for the incoherent gain of the Hanning window is also included). {\bf c}.) Superposition of parts {\bf a} and {\bf b}:  we see that the calculated shape closely matches the observed shape {\bf b}.
}
\end{figure}

\begin{figure}
\includegraphics[width=4in]{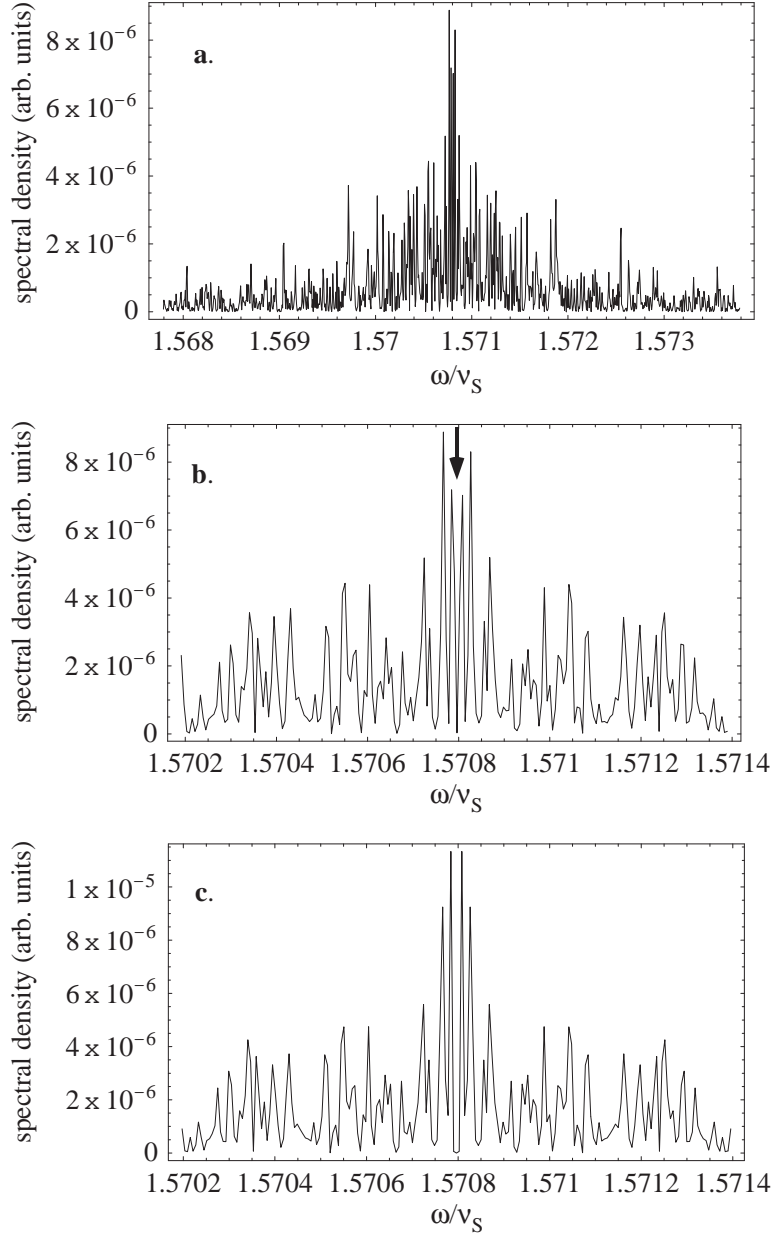}
\caption{\label{fig4} {\bf a}.) Spectrum of figure \ref{fig2}a zoomed and centered on the position of the small peak (i.e. $\omega/\nu_S = \pi/2$), with linear scales on both axes. {\bf b}.) An even closer view of the peak shows that it is actually a split peak: the arrow marks the position of the modulation frequency. {\bf c}.) This part shows what one obtains if one uses equation (\ref{obsPSD}) and the low-frequency part of the spectrum of figure \ref{fig2}a to calculate the expected shape of the peak. The calculation includes the correction for the incoherent gain of the Hanning window that has been used for trend removal. We see that the calculated shape {\bf c}. matches very well the observed shape {\bf b}.
}
\end{figure}

\begin{figure}
\includegraphics[width=4in]{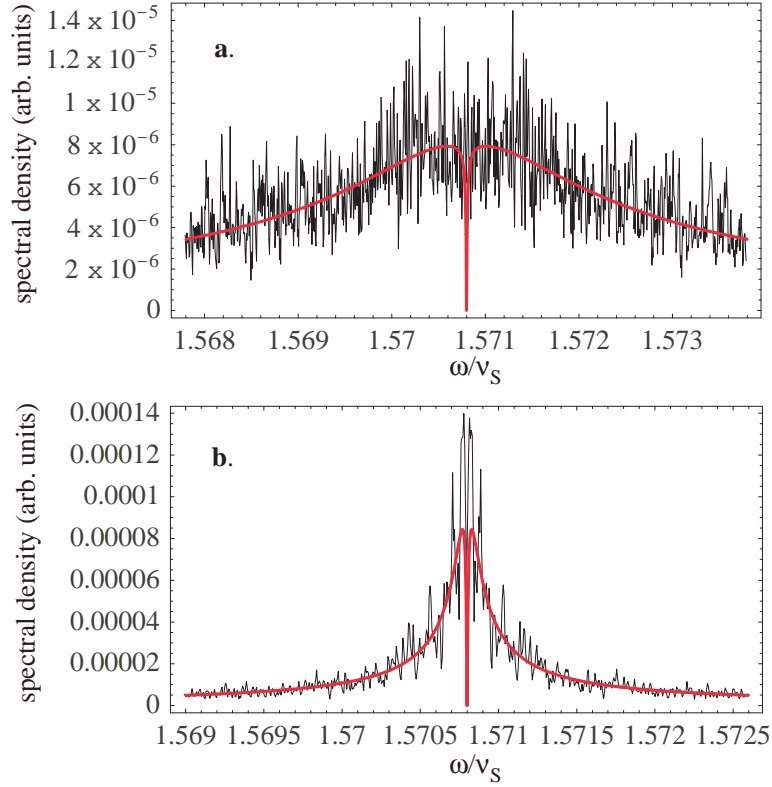}
\caption{\label{fig5} This figure illustrates how the peak splitting changes for different noise shapes: {\bf a}.) In this case the noise generation parameters are the same as for the signal in figure \ref{fig1}, except for the lowest relaxation rate $\lambda_{min} = 0.001$, and the figure shows the averaged spectrum zoomed and centered on the position of the small peak. The softer low-frequency behavior in noise spectrum produces side peaks that are smaller and further apart than those shown in figure \ref{fig3}. The solid line shows the expected behavior, calculated as in figure \ref{fig3}b {\bf b}.) A smaller value of the lowest relaxation rate, $\lambda_{min} = 0.00005$ yields instead much closer and higher side
peaks. Once again we see that the calculated shape (solid line) matches very well the observed shape. Because of the much narrower splitting, the scale in part {\bf b} is expanded with respect to {\bf a} to improve visibility.
}
\end{figure}

However equation (\ref{obsPSD}) also shows that the single peak -- in the case of true $1/f^\alpha$ noise -- or the side-peaks -- in the case of range-limited $1/f^\alpha$ noise --  can be modulated both in amplitude and in frequency by modulating either the noise spectrum or the relative timing shift amplitude, or the repetition index $m$: this means that uneven sampling of colored noise could be utilized to encode information, and since an encoding noise appears at first sight just ordinary noise, this could be used to implement a secure communication channel (there is a very rich literature on this topic, but here I give only a reference to a classic book \cite{GG} and to a recent paper \cite{cmaps}). A simple example of the kind of modulation that can be achieved can be gleaned from figures \ref{fig3},  \ref{fig4} and \ref{fig5}: if one uses the noise generator \cite{milotti,milotti2,milotti3}, it is possible to modulate the shape of the low-frequency part of the spectrum with a proper change of $\lambda_{min}$ and in this way one modulates in turn both the amplitude and the position, i.e., the frequency, of the side peaks.

\pagebreak

%figures

\end{document}